\documentstyle[11pt,newpasp,twoside]{article} 
\def\edcomment#1{\iffalse\marginpar{\raggedright\sl#1\/}\else\relax\fi} 
\reversemarginpar 

\begin{document} 
\title{A Photometric Method for estimating CNO Abundances in Globular Clusters}

\author{David Peat} 
\affil{Dept. of Physics \& Astronomy, University of Leeds, Leeds, UK.}
\author{Raymond Butler} 
\affil{Dept. of Physics, National University of Ireland, University Road, Galway, Ireland}

\section{Introduction} 

The estimation of variations in CNO abundances within globular
clusters yields information on the evolutionary history of the
clusters, and possibly also on the physical processes involved in
their formation.  This paper describes some preliminary results from a
new photometric method to estimate these variations. The method can be
used for any globular cluster, but M22 was chosen for a first
investigation because it is known (Richter, Hilker, \& Richtler 1999) to exhibit
large internal variations in CN absorption strength, and these
variations, unlike in most clusters, are believed (Anthony-Twarog,
Twarog, \& Craig 1995) to be {\it positively} correlated with CH
variations.

\section{Details of the Method}

The method uses the Stromgren indices v and b, the broad band index I,
together with a new index, defined as p, which consists of the shorter
wavelength half of the Stromgren v band.  The v band includes the CN
absorption band at 4215A, but the p band does not.  The index 2(v-p),
at a given I magnitude, is thus a measure of the CN absorption, and is
independent of differential reddening effects.  The colour index (p-I)
provides a suitably wide wavelength baseline for estimates of colour
temperature, and contains only little CN or CH absorption.  A
colour-magnitude diagram of I/(p-I) can be plotted, and a ``p anomaly"
can be defined as the displacement of an individual star parallel to
the (p-I) axis from the mean CMD.  A ``b anomaly'' is defined in the
same way, using (b-I). Correlations between the CN index and the p
anomalies can then be sought.

The evolutionary track of a star on the CMD depends on the molecular
weight of the material of the star, and this depends primarily on the
CNO abundances, together with a smaller dependence on the metal
abundances.  The "p anomaly" is thus an overall measure, through the
molecular weight, of variations in the CNO abundances.  Since the p
and I bands contain little CN or CH absorption, any correlation
between CN absorption and p anomaly would be suggestive of {\it
primordial} CNO variations, which have affected the subsequent
evolutionary tracks through the molecular weight.  Estimated
abundances, defined as del(log Z) where Z is the usual mean abundance
parameter for the cluster, can be derived from the p and b anomalies
for each star observed, using theoretical isochrones for different
chemical compositions.  Relative CN abundances can be derived from the
CN index, on the assumption that the logarithmic index is proportional
to the product of the C and N abundances at a given value of I.
Assuming the oxygen and metal abundances, it is then possible from the
two observations of CN and del(log Z), to estimate the two unknowns,
the C and N abundances.

\section{Results}

Results derived for a preliminary group, consisting of 26 giants and
29 subgiants in M22 were as follows:

(1)  There are substantial variations of both CN strengths and p
anomalies within the cluster, suggesting large variations of C and N
abundances.

(2) There seem to be three possible groups of stars, each group
showing a positive correlation between CN strength and p anomaly.  The
simplest explanation is that within each group there are carbon
abundance variations, with increasing carbon abundance increasing both
the CN strength and the molecular weight.  This would favour a
primordial origin for the abundance variations.

(3)   The differences between the three groups are, we suggest, due to
nitrogen abundance variations. However, the "strong CN" stars show p
anomalies broadly comparable to those of the "weak CN" stars,
suggesting that, for the molecular weights, increased nitrogen
abundances are offset by decreased carbon abundances.

(4)   Estimates of the relative nitrogen abundances in the three
groups were made from simple isochrones and the spectroscopic
assumption that the CN index is approximately proportional to the
logarithm of the product of the C and N abundances. Results suggested
log(N3/N1) = 1.3 \& log(N2/N1) = 0.7, with errors 0.3 and 0.2
respectively, for the subgiants; and log(N3/N1) = 0.9 \&
log(N2/N1) = 0.5, with similar errors, for the giant stars.  The
errors arise from uncertainties in the evolutionary and spectroscopic
models, as well as from uncertainties in the oxygen abundances which
affect the molecular weights.

(5)    It is possible that the relative nitrogen abundances in the
three groups can be expressed in the form 1:k:k$^2$, with k
approximately 4 for the subgiants, and 3 for the giants. These
relative abundances might be explained by postulating two processes
of star formation after an initial burst - with CNO cycling
synthesising nitrogen, subsequent ejection in supernovae
outbursts, followed by star formation from the same material in a
proto-cluster - each process increasing the nitrogen abundance by a
factor k.

       Further work will consist of analysis of a larger number of
stars, and also of analysis of a photometric band measuring the
strength of CH molecular absorption.

\end{document}